\documentclass[10pt,conference]{IEEEtran}
\IEEEoverridecommandlockouts
\usepackage{amsmath,amssymb,amsfonts}
\usepackage{algorithmic}
\usepackage{graphicx}
\usepackage{xcolor}
\usepackage{amsmath,lipsum}
\usepackage[ruled]{algorithm2e}
\usepackage{float}
\usepackage{subfigure}
\usepackage{multirow}
\usepackage{multicol}
\usepackage{cite}
\usepackage{stfloats}
\usepackage{booktabs}
\usepackage{caption}
\usepackage{textcomp}
\usepackage{arydshln}
\usepackage{makecell}
\usepackage{bm}
\usepackage[explicit]{titlesec}

\def\BibTeX{{\rm B\kern-.05em{\sc i\kern-.025em b}\kern-.08em
    T\kern-.1667em\lower.7ex\hbox{E}\kern-.125emX}}

\begin{document}

\title{Goal-Oriented Semantic Communication for Wireless Image Transmission via Stable Diffusion \\ }
\author{\IEEEauthorblockN{Nan~Li, Yansha~Deng }
\IEEEauthorblockA{Department of Engineering, King's College London.\\
Email: \{nan.3.li, yansha.deng\}@kcl.ac.uk}

\thanks{This work was supported in part by the Engineering and Physical Sciences Research Council (EPSRC), U.K., under Grant EP/W004348/1; and in part by UK Research and Innovation (UKRI) under the UK government’s Horizon Europe funding guarantee (grant number 10087666), as part of the European Commission-funded collaborative project MYRTUS, under the Smart Networks and Services Joint Undertaking (SNS JU) program (grant number 101135183).}

}
\maketitle

\begin{abstract}
Efficient image transmission is essential for seamless communication and collaboration within the visually-driven digital landscape. To achieve low latency and high-quality image reconstruction over a bandwidth-constrained noisy wireless channel, we propose a stable diffusion (SD)-based goal-oriented semantic communication (GSC) framework. In this framework, we design a semantic autoencoder that effectively extracts semantic information (SI) from images to reduce the transmission data size while ensuring high-quality reconstruction. Recognizing the impact of wireless channel noise on SI transmission, we propose an SD-based denoiser for GSC (SD-GSC) conditional on an instantaneous channel gain to remove the channel noise from the received noisy SI under known channel. For scenarios with unknown channel, we further propose a parallel SD denoiser for GSC (PSD-GSC)  to jointly learn the distribution of channel gains and denoise the received SI. It is shown that, with the known channel, our SD-GSC outperforms state-of-the-art ADJSCC and Latent-Diff DNSC, improving Peak Signal-to-Noise Ratio (PSNR) by 32\% and 21\%, and reducing Fréchet Inception Distance (FID) by 40\% and 35\%, respectively. With the unknown channel, our PSD-GSC improves PSNR by 8\% and reduces FID by 17\% compared to MMSE equalizer-enhanced SD-GSC. 

\end{abstract}

\begin{IEEEkeywords}
 Goal-oriented semantic communication, stable diffusion model, generative AI, and image transmission. 
\end{IEEEkeywords}

\begin{figure*}
    \centering
    \includegraphics[width =0.9 \textwidth]{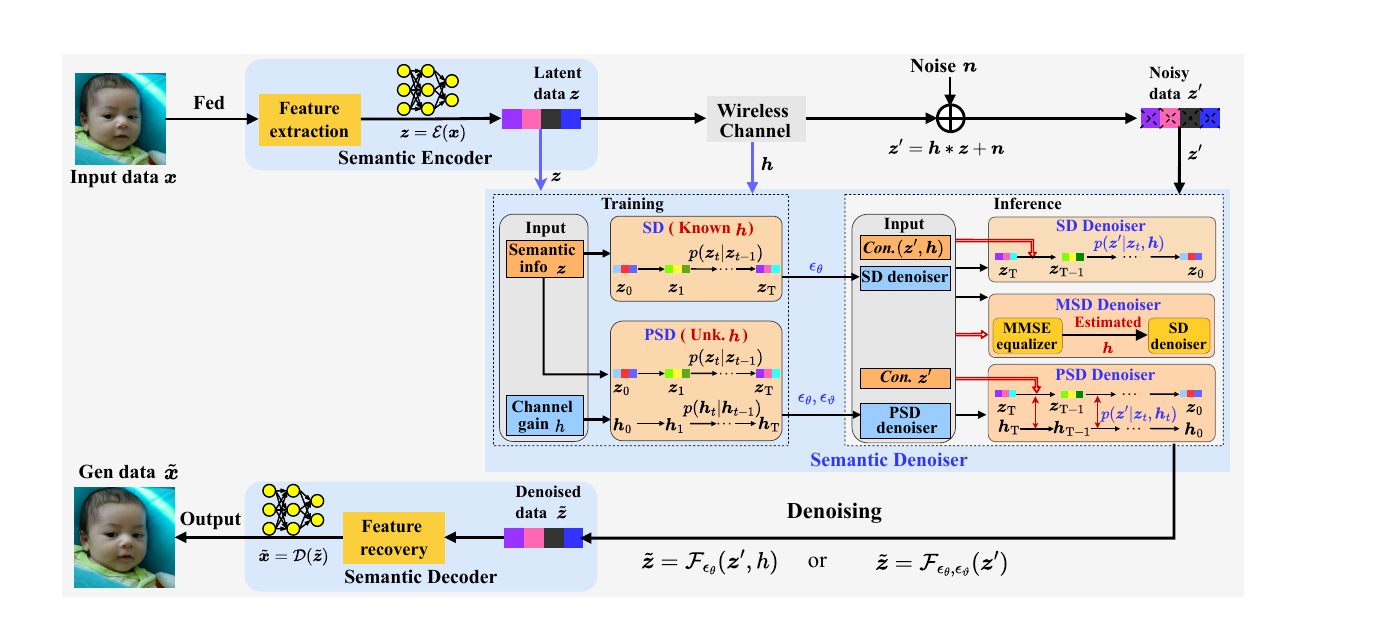}
    \caption{ Stable diffusion-based GSC framework for wireless image transmission}
    \label{fig:system_model}
\end{figure*}

\section{Introduction}
The growing reliance on smart visual applications, has dramatically driven the demand for seamless, high-quality image transmission.  Traditional communication systems that transmit full bit streams based on the Shannon's  technical framework, are struggling to meet these demands within bandwidth constraints. \textit{Semantic communication} (SC) has emerged as a promising approach to reduce redundancy and irrelevant information, leading to more efficient image transmission \cite{Luo_2022_WC}. 

To better capture task-specific semantic information (SI), \textit{goal-oriented semantic communication} (GSC) was introduced to incorporate both semantic and task-specific effectiveness level \cite{zhou2022task}. A joint source-channel coding (JSCC) scheme based on deep neural networks (DNN) was first developed in \cite{Bourtsoulatze_2019_TCCN} for high-resolution image transmission over fading channels. To better evaluate the semantic importance, \cite{Nan2024TWC,Nan2023Globecom} explored the attention mechanism to allocate varying attention weights to different features. However, the end-to-end design of source-channel coding in these schemes has limitations in adaptability, as end-to-end training needs to be performed for each task/goal with the lack of plug-and-play functionality of each module.

It is also important to note that the aforementioned works shared a common assumption of channel consistency between training and inference, breaking this assumption may cause significant performance degradation under dynamic conditions. To address this, \cite{Xu_2022_TCSVT} proposed an attention mechanism-based deep JSCC (ADJSCC) scheme that dynamically adjusts the signal-to-noise ratio (SNR) during training to adapt to fluctuated channel conditions. However, this scheme remains an end-to-end design of source- channel coding, limiting its flexibility and adaptation. Recent advances in generative AI, e.g., denoising diffusion probabilistic models (DDPM) \cite{chung2023diffusion}, offer new possibilities to address the adaptive denoising challenge in image generation \cite{Rombach_2022_CVPR}. Inspired by DDPM, \cite{Xu_2023_Globecom} proposed a novel plug-and-play module, Latent diffusion denoising SC (Latent-Diff DNSC), to characterize noisy channels and remove channel noise for image transmission. However, the uncontrollable generation process of DDPM makes this scheme challenging to perfectly reconstruct the image details like appearance, style and color \cite{Rombach_2022_CVPR}. Moreover, this scheme assumed perfect channel estimation, which poses challenges in practical scenarios with imperfect channel estimation.

To fill the gaps, we propose a stable diffusion (SD) model-based GSC framework for efficient image transmission over wireless fading channels, with the goal of achieving high-quality image reconstruction. Our main contributions are:
\begin{itemize}
    \item We propose an SD-based GSC framework with a semantic encoder at the transmitter to extract relevant SI, and a semantic denoiser and a semantic decoder at the receiver to denoise the received noisy SI and reconstruct the image, respectively. 
    \item For a known channel, we introduce an SD-based semantic denoiser for GSC (SD-GSC) that treats instantaneous channel gain as an input condition of SD to identify and remove noise. For an unknown channel, we propose a parallel SD-based semantic denoiser for GSC (PSD-GSC) that jointly estimates the instantaneous channel gain and denoises the received noisy SI.
    \item We conduct extensive experiments to examine the Peak SNR (PSNR) and Fréchet Inception Distance (FID). Under known channel conditions, SD-GSC demonstrates significant improvements compared to state-of-the-art ADJSCC \cite{Xu_2022_TCSVT} and Latent-Diff DNSC \cite{Xu_2023_Globecom}, improving PSNR by 32\% and 21\%, and reducing FID by 40\% and 35\%, respectively. Under unknown channel conditions, PSD-GSC improves PSNR by 8\% and reduces the FID by 17\% compared to MMSE equalizer-enhanced SD-GSC.  
\end{itemize}

\section{System Model and Problem Formulation}\label{section:system_model}
We consider an image transmission task between a transmitter and a receiver under Rayleigh fading channels. With the communication goal of achieving high-quality image
transmission, we introduce the SD-based GSC framework that comprises a semantic encoder at the transmitter, coupled with a semantic denoiser and semantic decoder at the receiver, as illustrated in Fig. \ref{fig:system_model}. The semantic encoder extracts the semantic information (SI) from the input image; the semantic decoder reconstructs the image from the received SI accordingly; the semantic denoiser removes wireless channel noise from the received SI to enhance the reconstruction capacity of the semantic decoder. Specifically, for known channel conditions where the channel gain is available, we propose an \textit{SD denoiser} to remove the noise while preserving the underlying SI in the latent vectors. For scenarios with unknown channel conditions, we introduce two following denoisers:
\begin{itemize}
    \item MMSE equalizer-enhanced SD (MSD) denoiser: An MMSE equalizer estimates the instantaneous channel gain, which is then fed into the SD denoiser.
    \item Parallel SD (PSD) denoiser: Two parallel SD modules jointly learn the instantaneous channel gain and denoise the channel noise.
\end{itemize}
The details of each module are presented one by one below.

At the transmitter, the input image  $\bm{x} \in \mathbb{R}^{\mathrm{H} \times \mathrm{W} \times \mathrm{C}}$, is first fed into the semantic encoder to extract the most critical and semantically relevant information related to the communication goal and outputs a lower-dimensional latent tensor $\bm{z} \in \mathbb{R}^{ \mathrm{d}}$, with $\mathrm{d} \ll \mathrm{H} \times \mathrm{W} \times \mathrm{C}$. Mathematically, we express this SI extraction process as 
\begin{equation}
    \bm{z} = \mathcal{E} (\bm{x}),
\end{equation}
where $\mathcal{E}(\cdot)$ denotes the operation of SI extraction. 

The SI \(\bm{z}\) is then transmitted over a wireless channel under Rayleigh fading, experiencing transmission impairments such as distortion and noise. The received noisy semantic information is mathematically represented as
\begin{equation}
    \bm{\bm{z}'} = \bm{h} * \bm{z} + \bm{n},
\end{equation}
where \(\bm{h}\) represents the channel gain between the transmitter and receiver, and \(\bm{n} \sim \mathcal{CN}(0, \sigma^2)\) is the additive white Gaussian noise (AWGN) with noise power \(\sigma^2\).

At the receiver, the received noisy SI ${\bm{z}'}$ is progressively denoised by semantic denoiser to produce the denoised SI as
\begin{equation}
\tilde{\bm{z}} = \mathcal{F}({\bm{z}'}),
\end{equation}
where $\mathcal{F}(\cdot)$ is the denosing operation of the semantic denoiser. 

The denoised SI $\tilde{\bm{z}}$ is then fed into the semantic decoder to generate the recovered image data $\tilde{ \textbf{x}} \in \mathbb{R}^{\mathrm{H} \times \mathrm{W} \times \mathrm{C}}$ as 
\begin{equation}
   \tilde{ \bm{x}} =  \mathcal{D}(\tilde{\bm{z}}),
\end{equation}
where $\mathcal{D}(\cdot)$ is the operation of the semantic encoder.

With the aim to accurately reconstruct the original image at the receiver while minimizing the distortion introduced by the wireless channel, we introduce the mean squared error (MSE) to measure the difference between the transmitted image at the transmitter and the recovered image at the receiver.
The objective is to minimize the average MSE between the original images and the reconstructed images as
\begin{equation}
    \min_{\mathcal{P}} \frac{1}{N} \sum_{\bm{x} \in \bf{X}} \big|\big|\bm{x} - \tilde{\bm{x}}\big|\big|^2,
\label{eq:mesloss}
\end{equation}
where $N$ is the size of dataset $\bf{X}$; $\mathcal{P}$ represents the set of learnable parameters in the various modules.

\section{Architecture of SD-based GSC}
In this section, we present the details of the semantic encoder and decoder, and the semantic denoisers under known and unknown channels, respectively.

\subsection{Semantic Encoder and Semantic Decoder}
To extract the relevant SI while minimizing the transmitted data size and optimizing the quality of image reconstruction, we propose a DNN-based semantic encoder and decoder. The architecture follows a downsampling and upsampling framework similar to U-Net \cite{Unet}.

In the semantic encoder, the input  \(\bm{x}\) first passes through a convolutional layer \((3 \times 3 \, \text{kernel}, \text{stride} \, 2, \text{padding} \, 1)\), a residual block and an attention block, both operating on the RGB channels \((\mathrm{C}=3)\) to capture key features. This is followed by three downsampling stages, each comprising a convolutional layer \((3 \times 3 \, \text{kernel}, \text{stride} \, 2, \text{padding} \, 1)\), a residual block, and an attention block, progressively halving the spatial dimensions while increasing the channel dimensions \((c \in \{64, 128, 256, 512\})\) to capture more abstract features. Batch normalization and ReLU activation are then applied, and the output is embedded into a latent representation \(\bm{z} \in \mathbb{R}^{ \mathrm{d}}\) via a linear layer. The semantic decoder transforms the denoised SI \(\tilde{\bm{z}}\) into a higher-dimensional feature space \( \mathbb{R}^{\mathrm{h} \times \mathrm{w} \times \mathrm{c}}\) using a linear layer. It then performs upsampling in three stages, each with a transposed convolutional layer \((3 \times 3 \, \text{kernel}, \text{stride} \, 2, \text{padding} \, 1)\) to double the spatial dimensions, followed by a residual block and an attention block to refine features and highlight important spatial details. As these stages progress, the channel dimensions reduce \((c \in \{512, 256, 128, 64\})\), expanding the spatial dimensions. The final transposed convolutional layer \((3 \times 3 \, \text{kernel}, \text{stride} \, 1, \text{padding} \, 1)\) and a sigmoid activation function produce the reconstructed image \(\tilde{\bm{x}} \in \mathbb{R}^{\mathrm{H} \times \mathrm{W} \times \mathrm{C}}\), normalized to [0, 1], with denormalization scaling the output to [0, 255] for accurate pixel values.

\subsection{SD Denoiser under Known Channel}
To handle the wireless channel noise during SI transmission, the uncontrolled image generation process of DDPM \cite{NEURIPS2020_4c5bcfec}, as applied in Latent-Diff DNSC \cite{Xu_2023_Globecom}, potentially degrades the reconstruction quality \cite{Rombach_2022_CVPR}. To overcome this challenge, we propose the \textit{SD denoiser} that uses the received noisy SI and instantaneous channel gain as control conditions to effectively remove channel noise, as shown in Fig. \ref{fig:system_model}.

\subsubsection{Forward Process of DDPM}
The forward process (i.e., training process) is applied to the transmitted SI, \(\bm{z}\), generated by the semantic encoder. This involves iteratively adding Gaussian noise to the initial distribution \(\bm{z}_0 \sim p(\bm{z})\) over \(\mathrm{T}\) time steps, gradually approaching an isotropic Gaussian distribution \(\bm{z}_\mathrm{T} \sim \mathcal{N}(0, \mathbf{I})\). At time step $t \in [0, \mathrm{T}] $, the forward process of $\bm{z}_t $ is expressed as 
\begin{equation}
\bm{z}_{t}=\sqrt{1-\beta_{t}} \bm{z}_{t-1}+\sqrt{\beta_{t}} \epsilon,
\label{eq:forward}
\end{equation}
where $\beta_{t} \in (0,1)$ is the noise scheduling function, typically modeled as a monotonically increasing linear function of $t$, and $\epsilon \sim \mathcal{N}(0, \mathbf{I})$. 

From the score-based perspective (i.e., the gradient of the log probability density with respect to the data $\bm{z}_t$ at each noise scale $\beta_t$), the forward process can be expressed as the following linear stochastic differentiable equation (SDE) 
\begin{equation}
    \mathrm{d}^{(f)} \bm{z}_t = - \frac{\beta_t}{2} \bm{z}_t \mathrm{d}t + \sqrt{\beta_t} \mathrm{d} \bm{w},
    \label{eq:addnoise}
\end{equation}
where $\bm{w}$ is the $d$-dimensional Wiener process. 

\subsubsection{Reverse Process of DDPM}
The reverse process (i.e., inference process) of DDPM aims to recover the original SI \(\bm{z}\) from the noisy sample $\bm{z}_\mathrm{T} \sim \mathcal{N}(0, \mathbf{I})$. This process mirrors the marginal distribution of the forward process \(p(\bm{z}_t)\) but with the drift direction reversed. According to (\ref{eq:addnoise}), the reverse process of DDPM can be achieved by the corresponding reverse SDE as \cite{chung2023diffusion}
\begin{equation}
    \mathrm{d}^{(r)} \bm{z}_t = \big [ - \frac{\beta_t}{2} \bm{z}_t - \beta_t \nabla_{\bm{z}_t} \log p (\bm{z}_t) \big ] \mathrm{d} t + \sqrt{\beta_t} \mathrm{d} \overline{\bm{w}},
    \label{eq:reverse}
\end{equation}
where $\mathrm{d} t$ is the reverse time step, $\mathrm{d}\overline{w}_t$ is the the standard Wiener process in reverse. 

Since $\nabla_{\bm{z}_t} \log p (\bm{z}_t) = \nabla_{\boldsymbol{\bm{z}}_t} \log p\left(\boldsymbol{\bm{z}}_t \mid \boldsymbol{\bm{z}}_0\right)$, we can approximate $\nabla_{\bm{z}_t} \log p (\bm{z}_t) \simeq	 \boldsymbol{s}_\theta\left(\boldsymbol{\bm{z}}_t, t\right)$ for the reverse process in (\ref{eq:reverse}) by solving the following minimization problem during the training in the forward process \cite{Vincent2011NC}: \begin{equation}
\theta^* = \underset{\theta}{\operatorname{argmin}} \mathbb{E}_{\bm{z}_t, \bm{z}_0}\left[\left\|\boldsymbol{s}_\theta\left(\boldsymbol{\bm{z}}_t, t\right)-\nabla_{\boldsymbol{\bm{z}}_t} \log p\left(\boldsymbol{\bm{z}}_t | \boldsymbol{\bm{z}}_0\right)\right\|_2^2\right],
\label{eq:trainprocess}
\end{equation}
where the trained score network $\bm{s}_\theta\left(\bm{z}_t, t\right) $ can be denoted by using Tweedie’s identity \cite{NEURIPS2021_077b83af} as
\begin{equation}
    \bm{s}_\theta\left(\boldsymbol{\bm{z}}_t, t\right) = \nabla_{\bm{z}_t} \log p(\bm{z}_t)  = - \frac{1}{\sqrt{1-\bar{\alpha}_{t}}}\epsilon_\theta\left(\boldsymbol{\bm{z}}_t, t\right),
    \label{eq:score}
\end{equation}
where $\alpha_t = 1-\beta_t$ and $\bar{\alpha}_{t}=\prod_{i=1}^{t}\left(1-\alpha_{i}\right)$, and the parameter $\epsilon_\theta\left(\boldsymbol{\bm{z}}_t, t\right)$ is the learned noise estimator at time step $t$.

\subsubsection{Reverse Process of Stable Diffusion}
Evidently, the reverse process of DDPM from a random Gaussian sample $\bm{z}_\mathrm{T}$ cannot ensure the reconstruction quality. To allow for more controllable and guided image generation, we design an \textit{SD denoiser} conditional on the received noisy SI ${\bm{z}'}$ and instantaneous channel gain ${\bm{h}}$. Here, the \textit{SD denoiser} follows the same forward process as DDPM \cite{Rombach_2022_CVPR} .

Leveraging the diffusion model as the prior, it is straightforward to modify (\ref{eq:reverse}) to derive the reverse process of SD from the posterior distribution as
\begin{equation}
\begin{aligned}
   \mathrm{d}^{(r)} \bm{z}_t & = [ - \frac{\beta_{t}}{2} \bm{z}_t - \beta_{t} \nabla_{\bm{z}_t} \log p\left(\bm{z}' \mid \bm{z}_t\right) ] \mathrm{d} t + \sqrt{\beta_{t}} \mathrm{d} \overline{\bm{w}} \\ 
   & = [ - \frac{\beta_{t}}{2} \bm{z}_t - \beta_{t} (\nabla_{\bm{z}_t} \log p (\bm{z}_t)  \\
   & \hspace{5mm}+ \nabla_{\bm{z}_t} \log p(\bm{z}'|\bm{z}_t ,\bm{h}) )] \mathrm{d} t +\sqrt{\beta_{t}} \mathrm{d} \overline{\bm{w}}.
    \label{eq:con_reverse}
\end{aligned}
\vspace{-1mm}
\end{equation}

By discretizing the reverse process in (\ref{eq:con_reverse}), we have
\begin{equation}
\begin{aligned}
    \bm{z}_{t-1} = & \frac{1}{\sqrt{\alpha_{t}} }(\bm{z}_{t} + \beta_t [\bm{s}_\theta\left(\boldsymbol{\bm{z}}_t, t\right) + \nabla_{\bm{z}_t} \log p(\bm{z}'|\bm{z}_t,\bm{h}) ] ) \\
    & + \sqrt{\beta_t} \mathcal{N}(0, \mathbf{I}),
\end{aligned}
\label{eq:conditionreverse}
\end{equation}
where $\bm{s}_\theta\left(\boldsymbol{\bm{z}}_t, t\right)$ is the trained score network in (\ref{eq:score}).

To solve the reverse process in (\ref{eq:conditionreverse}), the main challenge lies in the posterior distribution $p(\bm{z}'|\bm{z}_t,\bm{h})$. While the relationship between the received noisy SI \(\bm{z}'\) and the transmitted SI \(\bm{z}_0\) is known, the relationship between the intermediate data \(\bm{z}_t\) at the \(t\)th step of the forward process and \(\bm{z}'\) remains unknown. To tackle this issue, we express $p\left(\bm{z}' \mid \bm{z}_t\right)$ as 
\begin{equation}
    p\left(\bm{z}' \mid \bm{z}_t\right)=\int p\left(\bm{z}' \mid \bm{z}_0 \right) p\left(\bm{z}_0 \mid \bm{z}_t\right) \mathrm{d} \bm{z}_0.
    \label{eq:marginal}
\end{equation}
where the mean of $p(\bm{z}_0\mid \bm{z}_t)$ can be approximated by a delta function as
\begin{equation}
     p\left(\bm{z}_0 \mid \bm{z}_t\right) \simeq	 \delta_{\textbf{E}[\bm{z}_0 \mid \bm{z}_t]}(\bm{z}_0),
\end{equation}

To estimate the $E[\bm{z}_0|\bm{z}_t]$, we can use the well-trained noise estimator $\epsilon_{\theta}(\bm{z}_t, t)$ in the forward process (\ref{eq:trainprocess}) to obtain the estimation $ E[\bm{z}_0|\bm{z}_t] = \hat{\bm{z}}_t$ as
\begin{equation}
    \hat{\bm{z}}_t =\frac{1}{\sqrt{\alpha_{t}} }(\bm{z}_{t}- \sqrt{1-\bar{\alpha}_{t-1}}\epsilon_{\theta}(\bm{z}_t,t))
    \label{eq:z0}.
\end{equation}

Using (\ref{eq:z0}), the approximation $p(\bm{z}_0|\bm{z}_t)$ leads to the following formula for the gradient of the log-likelihood:
\begin{equation}
    \nabla_{\bm{z}_t} \log p(\bm{z}'|\bm{z}_t,\bm{h}) \hspace{-1mm}= \hspace{-1mm}- \frac{1-\bar{\alpha}_{t}}{\left(1 \hspace{-1mm}-\hspace{-1mm}\alpha_{t}\right)\left(1\hspace{-1mm}-\hspace{-1mm}\bar{\alpha}_{t-1}\right)} \nabla_{\bm{z}_t} ||\bm{z}'\hspace{-1mm}-\hspace{-1mm}\bm{h} * {\bm{z}_t}||^2.
\end{equation}
\subsection{PSD Denoiser under Unknown Channel}
Notably, the \textit{SD denoiser} is only applicable when the instantaneous channel gain $\bm{h}$ is known, and hence cannot be directly used for the scenarios with imperfect estimation of $\bm{h}$ (e.g., Massive MIMO communications systems). To solve this issue, we propose a parallel SD  (PSD) denoiser to jointly estimate the channel gain and remove the noise. 

\subsubsection{Forward Process}
Since $\bm{z}$ and $\bm{h}$ are independent, the posterior probability is given by 
\begin{equation}
    p(\bm{z},\bm{h}|\bm{\bm{z}'}) \propto p(\bm{\bm{z}'}|\bm{z},\bm{h}) p(\bm{z}) p(\bm{h}).
\end{equation}
Therefore, we can train two separate forward processes for $\bm{z}$ and $\bm{h}$, respectively. Similar to (\ref{eq:score}) and (\ref{eq:z0}), the score network of $\bm{h}$ and  $ E[\bm{h}_0|\bm{h}_t]= \hat{\bm{h}}_t$ can be estimated using
\begin{equation}
    \bm{s}_{\vartheta} \left(\boldsymbol{\bm{h}}_t, t\right) = \nabla_{\bm{h}_t} \log p(\bm{h}_t)  = - \frac{1}{\sqrt{1-\bar{\alpha}_{t}}}\epsilon_{\vartheta}(\bm{h}_t,t),
    \label{eq:score1}
\end{equation}
and
\begin{equation}
    \hat{\bm{h}}_t =\frac{1}{\sqrt{\alpha_{t}} }(\bm{h}_{t}- \sqrt{1-\bar{\alpha}_{t-1}}\epsilon_{\vartheta}(\bm{z}_t,t))
    \label{eq:h0},
\end{equation}
where $\epsilon_\vartheta\left(\boldsymbol{\bm{h}}_t, t\right)$ is the learned noise estimator of channel gain with parameter $\theta$ at time step $t$.
\subsubsection{Reverse Process}
Similar to the reverse process in (\ref{eq:con_reverse}), the two reverse processes of PSD can be expressed as
\vspace{-0.5mm}
\begin{equation}
\begin{aligned}
   \mathrm{d}^{(r)} \bm{z}_t = & \big[ - \frac{\beta_{t}}{2} \bm{z}_t - \beta_{t} (\nabla_{\bm{z}_t} \log p (\bm{z}_t) \\ 
   & + \nabla_{{\bm{z}}_t} \log p(\bm{z}'|\bm{z}_t, \bm{h}_t) )\big] \mathrm{d} t + \sqrt{\beta_{t}} \mathrm{d} \overline{\bm{w}},
\end{aligned}
\end{equation}
and
\vspace{-2mm}
\begin{equation}
\begin{aligned}
   \mathrm{d}^{(r)} \bm{h}_t = & \big[ - \frac{\beta_{t}}{2} \bm{h}_t - \beta_{t} (\nabla_{\bm{h}_t} \log p (\bm{h}_t) \\ 
   & + \nabla_{{\bm{h}}_t} \log p(\bm{z}'|\bm{z}_t, \bm{h}_t) ) \big] \mathrm{d} t + \sqrt{\beta_{t}} \mathrm{d} \overline{\bm{w}}.
\end{aligned}
\end{equation}
Correspondingly, the above two reverse processes can be discretized in a similar way to (\ref{eq:conditionreverse}) and expressed as 
\vspace{-1mm}
\begin{align}
    \bm{z}_{t-1} = & \frac{1}{\sqrt{\alpha_{t}} }\big(\bm{z}_{t} +  \beta_t \big[\nabla_{\bm{z}_t} \log p(\bm{z}'|\bm{z}_t,\bm{h}_t) +  \bm{s}_{\theta} \left(\boldsymbol{\bm{z}}_t, t\right) \big] \big)  
 \nonumber \\
    &  + \sqrt{\beta_t} \mathcal{N}(0, \mathbf{I}),
\end{align}
and 
\vspace{-3mm}
\begin{align}
    \bm{h}_{t-1} =  & \frac{1}{\sqrt{\alpha_{t}} }\big(\bm{h}_{t} +  \beta_t \big[\nabla_{\bm{h}_t} \log p(\bm{z}'|\bm{z}_t,\bm{h}_t)+  \bm{s}_{\vartheta} \left(\boldsymbol{\bm{h}}_t, t\right) \big] \big) \nonumber  \\
    &  + \sqrt{\beta_t} \mathcal{N}(0, \mathbf{I}).
\end{align}

Due to the sparse structure of wireless channel, we use $\ell_1 $ regularization to sparse the channel gain by augmenting the diffusion prior thereby 
better stabilize the reconstruction. The minimization strategy for the channel gain then becomes
\begin{equation}
    \bm{h}_{t-1} = \bm{h}_{t-1} - \alpha (||\bm{\bm{z}'}-\bm{h}_{t-1} * \bm{z}_{t-1}||_2 + \phi ||\bm{h}_{t-1}||),
\end{equation}
where $\phi$ is the regularization strength. 
\subsection{Training and Implementation}
The proposed SD-based GSC architecture can be effectively trained via an end-to-end approach; however, this method may suffer from slow convergence. To mitigate this issue, we employ a step-by-step training strategy that systematically trains each component of the SD-based GSC architecture.

\begin{algorithm}[t]
\caption{Training and Implementation}
\label{alg:semantic-denoiser}
\begin{algorithmic}[1]
\REQUIRE Training dataset $\bf{X}$, time steps $T$, hyperparameters $\left\{\beta_1, \cdots, \beta_T\right\}$, $\zeta_t^\theta$, $\zeta_t^\vartheta$, and $\phi$
\ENSURE Semantic denoisers $\zeta_t^\theta$, $\zeta_t^\vartheta$

\STATE Train the semantic encoder and decoder using (\ref{eq:aeloss}).

\tcp{\textbf{\textit{Training SD denoiser}}}
\STATE Freeze parameters of semantic extraction module $\mathcal{E}(\cdot)$.
\REPEAT
    \STATE Sample a latent representation $\bm{z}_0 \sim p(\bm{z})$.
    \STATE $t \sim \textrm{Uniform}(\{1,2,\ldots, T\})$, $\epsilon \sim \mathcal{N}(0, \mathbf{I})$.
    \STATE Take gradient descent step on $\nabla_{\theta} \left\|\epsilon_{\theta}(\bm{z}_t, t) - \epsilon \right\|_{2}^{2}$.
\UNTIL Converged

\tcp{\textbf{\textit{Denoising of SD denoiser}}}
\STATE Sample $\bm{z}_T \sim \mathcal{N}(0, \mathbf{I})$
\FOR{denoising step $t = T, \ldots, 1$}
    \STATE $\hat{\epsilon} \leftarrow \epsilon(\bm{z}_t, t)$.
    \STATE Estimate $\hat{\bm{z}}_t$ using (\ref{eq:z0}).
    \STATE $g \leftarrow \nabla_{\bm{z}_t} \log p(\bm{z}'|\bm{z}_t,\bm{h})$.
    \STATE Compute the conditional score $\bm{s} \leftarrow \zeta_t^\theta g - \frac{1}{\sqrt{1-\bar{\alpha}_{t}}} \hat{\epsilon}$. 
    \STATE Sample $\varphi \sim \mathcal{N}(0, \mathbf{I})$. 
    \STATE Compute $\bm{z}_{t-1} =  \frac{1}{\sqrt{\alpha_{t}} }(\bm{z}_{t} + \beta_t \bm{s}) + \sqrt{\beta_t} \varphi$.
\ENDFOR

\tcp{\textbf{\textit{Training PSD denoiser}}}
\STATE Train denoising estimators $\epsilon_\theta$ and $\epsilon_{\vartheta}$ for $\bm{z}$ and $\bm{h}$ following steps 3-7, respectively.

\tcp{\textbf{\textit{Denoising of PSD denoiser}}}
\STATE Sample $\bm{z}_T, \bm{h}_T \sim \mathcal{N}(0, \mathbf{I})$.
\FOR{denoising step $t = T, \ldots, 1$}
    \STATE $\hat{\epsilon}_\theta \leftarrow \epsilon_\theta(\bm{z}_t, t)$, $\hat{\epsilon}_\vartheta \leftarrow \epsilon_\vartheta (\bm{z}_t, t)$.
    \STATE Estimate $\hat{\bm{z}}_t$ and $\hat{\bm{h}}_t$ using (\ref{eq:z0}) and (\ref{eq:h0}).
    \STATE $g_\theta \leftarrow \nabla_{\bm{z}_t} \log p(\bm{z}'|\bm{z}_t,\bm{h}_t)$, $g_\vartheta \leftarrow \nabla_{\bm{h}_t} \log p(\bm{z}'|\bm{z}_t,\bm{h}_t)$.
    \STATE Compute the conditional score \\
        $\small \bm{s}_\theta \leftarrow\zeta_t ^\theta g_\theta - \frac{1}{\sqrt{1-\bar{\alpha}_{t}}} \hat{\epsilon}_\theta$ and $\bm{s}_\vartheta \leftarrow\zeta_t ^\vartheta g_\vartheta - \frac{1}{\sqrt{1-\bar{\alpha}_{t}}} \hat{\epsilon}_\vartheta$.
    \STATE Sample $\varphi_\theta, \varphi_\vartheta \sim \mathcal{N}(0, I)$.
    \STATE Compute $\bm{z}_{t-1} =  \frac{1}{\sqrt{\alpha_{t}} }(\bm{z}_{t} + \beta_t \bm{s}_\theta) + \sqrt{\beta_t} \varphi_\theta$ and $\bm{h}_{t-1} =  \frac{1}{\sqrt{\alpha_{t}} }(\bm{h}_{t} + \beta_t \bm{s}_\vartheta) + \sqrt{\beta_t} \varphi_\vartheta$.
    \STATE $\bm{h}_{t-1} \leftarrow \bm{h}_{t-1} - \alpha (||\bm{\bm{z}'} - \bm{h}_{t-1}  \bm{z}_{t-1}||_2 + \phi ||\bm{h}_{t-1}||)$.
\ENDFOR

\tcp{\textbf{Fine-tuning process}}

\STATE Freeze the parameters of the semantic encoder and semantic denoisers, then
fed the denoised SI $\tilde{\bm{z}}$ to finetune the semantic decoder using MSE loss in (\ref{eq:mesloss}).
\end{algorithmic}
\end{algorithm}

Initially, we start by jointly training the semantic encoder and semantic decoder by incorporating MSE and Kullback-Leibler (KL) divergence into the loss function to better reconstruct the image. The MSE facilitates accurate reconstruction, while the KL divergence serves as a regularization term, guiding the data distribution in the latent space towards approximating a unit Gaussian distribution. The loss function is defined as the sum of MSE and KL divergence via

\begin{equation}
    \mathcal{L} = \frac{1}{N} \sum_{\bm{x} \in \bf{X}} \big|\big|\bm{x} - \tilde{\bm{x}}\big|\big|^2  + \lambda \mathcal{L}_{\text{KL}},
    \label{eq:aeloss}
\end{equation}
where \(\lambda\) controls the significance of KL divergence relative to MSE. The KL divergence is expressed as
\begin{equation}
\mathcal{L}_{\text{KL}} = -\frac{1}{2} \sum_{i=1}^{N} (1 + \log(\sigma_i^2) - \mu_i^2 - \sigma_i^2),
\label{eq:kl}
\end{equation}
where \(\mu_i\) and \(\sigma_i\) denote the mean and standard deviation of the latent space, respectively.

Following this, we freeze the parameters of the semantic encoder and feed the latent semantic vector \( \bm{z} \) into the semantic denoiser for training the \textit{SD denoiser} and \textit{PSD denoiser} under both known and unknown channels. Once each component is individually trained, we finalize the process by fine-tuning the entire SD-based GSC architecture in an end-to-end manner. This fine-tuning stage optimizes the collaboration between all modules, resolving any discrepancies that may arise from independently trained components. The detailed training and implementation process is presented in \textbf{Algorithm 1}.

\section{Performance Evaluation}\label{Performance}

We consider the end-to-end transmission of image tasks from a device (i.e., NVIDIA RTX 2080TI) to a server (i.e., NVIDIA A100 80GB) over a Rayleigh fading channel. The FFHQ dataset \cite{Karras_2019_CVPR}, consists of 70,000 high-quality human face images at a resolution of 256×256 pixels, is used as the benchmark for this evaluation. The implementation of SD-based GSC framework is built on Ubuntu 22.04 using PyTorch. For the semantic denoiser, we set the number of time steps \(\mathrm{T} = 1000\) and employ a linear variance scheme to determine the hyperparameters \(\{\beta_1, \beta_2, \ldots, \beta_\mathrm{T}\}\). The learning rate is initialized at 0.001, and stochastic gradient descent (SGD) is used as the optimizer for the loss function.

\textbf{Baselines:} To demonstrate the effectiveness of our proposed frame, we perform a comparative analysis against the state-of-the-art baselines: ADJSCC \cite{Xu_2022_TCSVT} and Latent-Diff DNSC \cite{Xu_2023_Globecom}. We also compare with MSD denoiser-based GSC (MSD-GSC) as a benchmark for channel gain estimation to highlight the effectiveness of our PSD-GSC in learning channel gains. 

\textbf{Metrics:} We evaluate two performances: peak signal-to-noise ratio (PSNR) to evaluate image reconstruction quality, with higher values indicating better reconstruction quality; and Frechet inception distance (FID) to assess the similarity between real and reconstructed image distributions, with lower values indicating better reconstruction quality. 
\begin{figure}
    \centering
    \subfigure[PSNR]{
    \includegraphics[width=0.33 \textwidth]{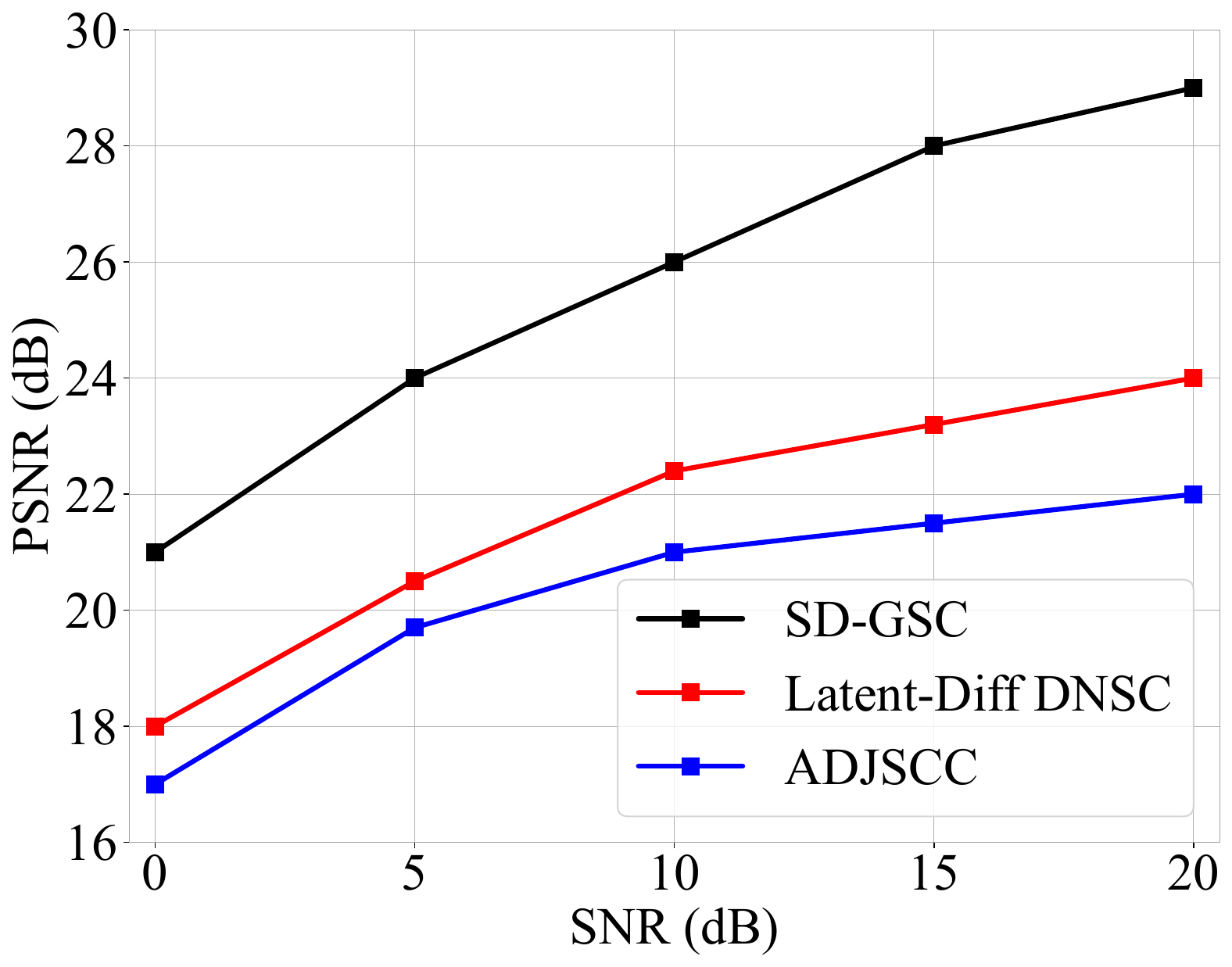}}
    \subfigure[FID]{
    \includegraphics[width=0.33 \textwidth]{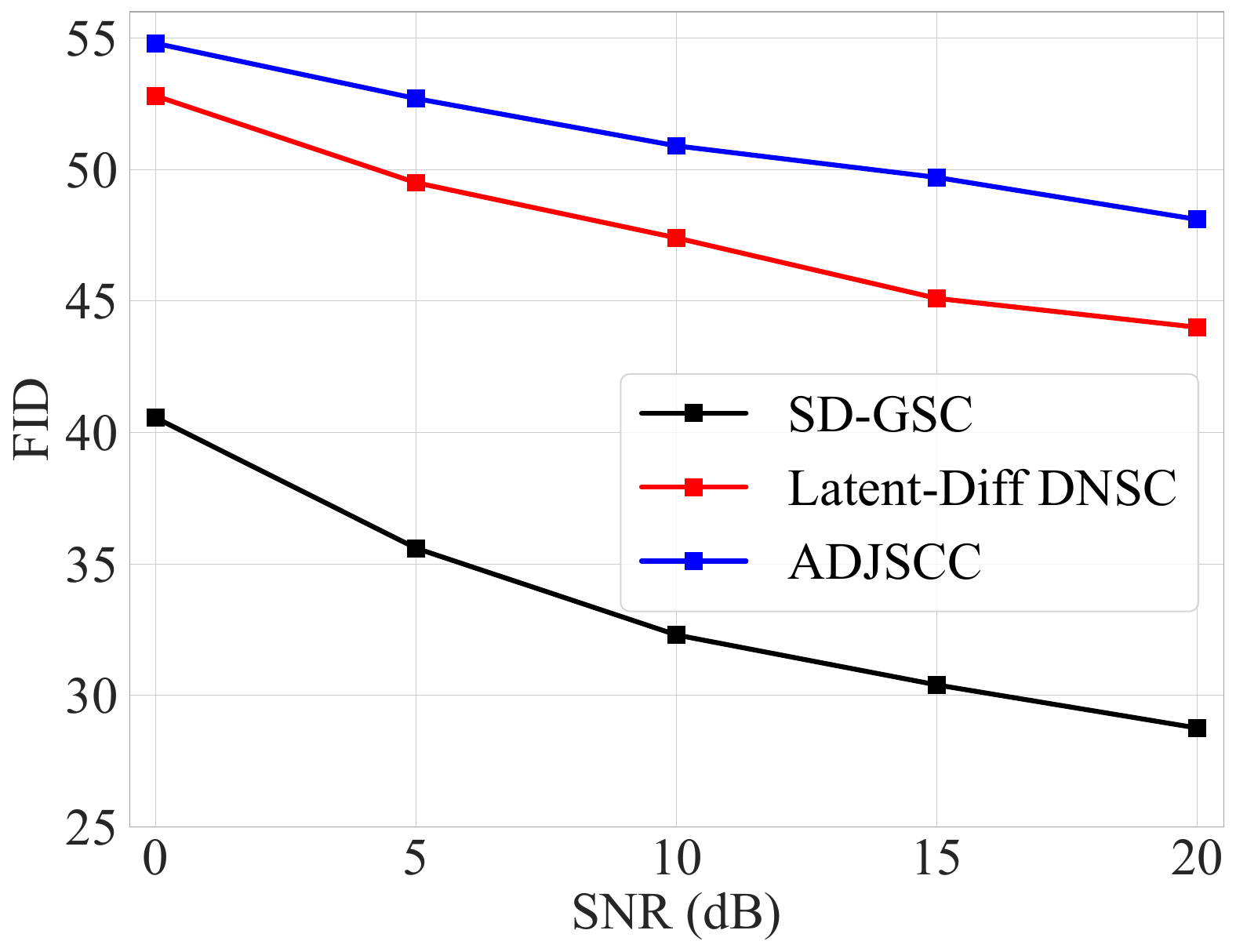}}
    \caption{PSNR and FID under known channel}
    \label{fig:knownH}
\end{figure}

\subsection{Known Channel}
Fig. \ref{fig:knownH} evaluates the PSNR and the FID performance across various SNRs under known channel. As expected, all the metrics improve as the SNR increases, since better wireless channel conditions facilitate higher-quality image reconstruction. Interestingly, SD-GSC and Latent-Diff DNSC outperform ADJSCC in both metrics, due to the diffusion model's ability in capturing wireless channel characteristics and removing noise effectively, resulting in more accurate reconstruction. Notably, SD-GSC outperforms two baselines in both metrics. Specifically, SD-GSC achieves approximately  32\% and 21\% PSNR improvement, and 40\% and 35\% FID reduction as compared to ADJSCC and Latent-Diff DNSC, respectively. This superior performance arises from the integration of the instantaneous channel gain, which enables better characterization of channel gain and mitigation of distortion caused by wirless channel noise, resulting in a more controlled and guided image generation than the uncontrolled generation process of DDPM used in Latent-Diff DNSC.

\subsection{Unkown Channel}
\begin{figure}
    \centering
    \subfigure[PSNR]{
    \includegraphics[width=0.325\textwidth]{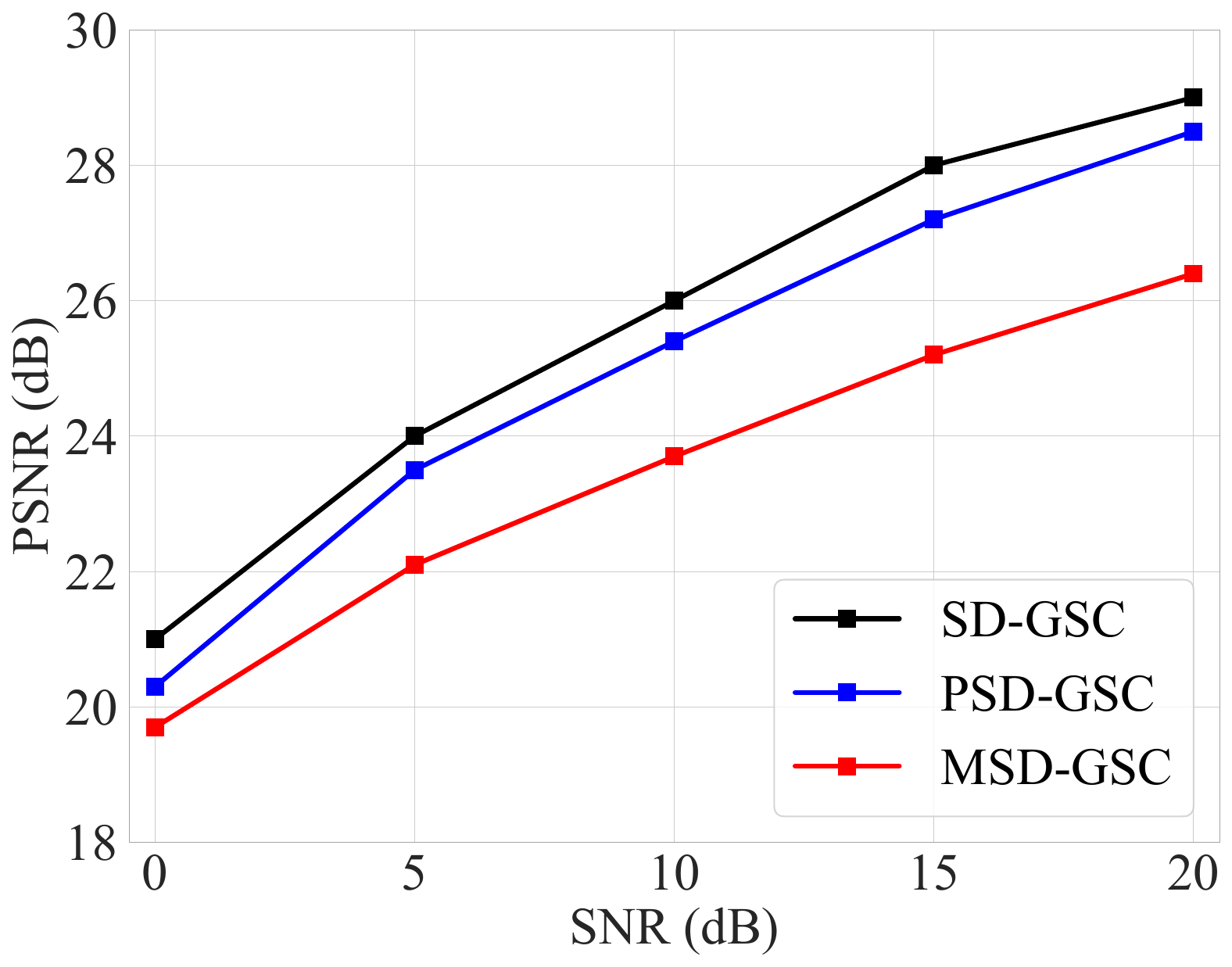}}
    \subfigure[FID]{
    \includegraphics[width=0.325 \textwidth]{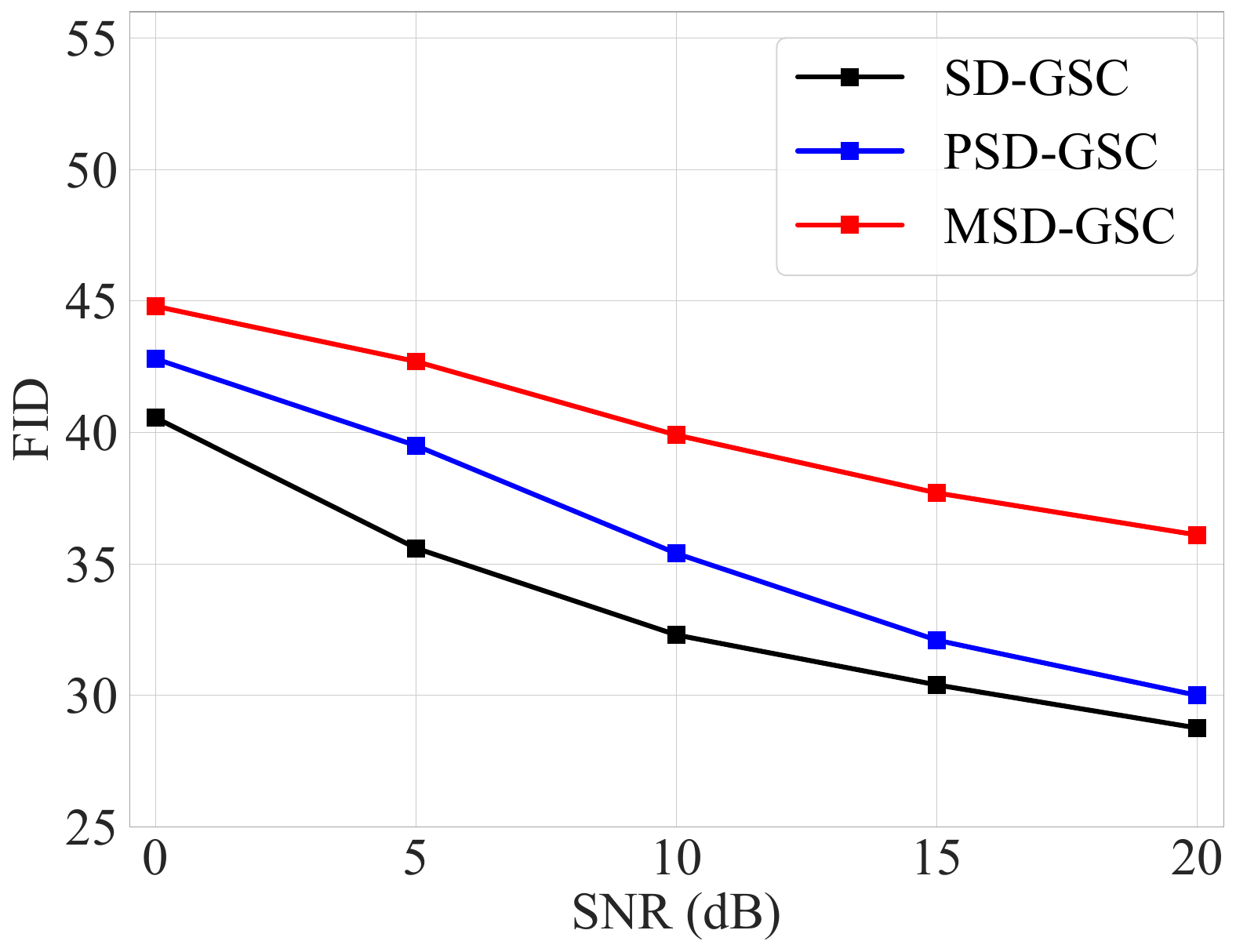}}
    \caption{PSNR and FID under unknown channel}
    \label{fig:unknownH}
\end{figure}

To validate the effectiveness of our proposed PSD-GSC for channel estimation as well as efficient image transmission, we also compare our proposed PSD-GSC with MSD-GSC and SD-GSC.  
Fig. \ref{fig:unknownH} presents the PSNR and FID performance for various SNRs under unknown channel. We can see both PSD-GSC and MSD-GSC experience degradation in both metrics compared to SD-GSC with known channel gain $h$. Specifically, our proposed PSD-GSC and MSD-GSC experience degradation in PSNR by approximately 1.7\% and 9.0\%, and improvement in FID by 4.3\% and 25.5\%, respectively. This is because channel estimation errors result in inaccurate channel gain inputs of the diffusion model, which decreases its denoising and high-quality image reconstruction capabilities. Importantly, PSD-GSC outperforms MSD-GSC in both metrics, improving PSNR by 8\% and reducing FID by 17\%, respectively. This is due to the probabilistic modeling-based diffusion model used in PSD-GSC can well estimate nonlinear channels and capture the complex properties of wireless channels, for better image reconstruction.

\section{Conclusion}\label{conclusion}
In this paper, we developed a stable diffusion-based goal-oriented semantic communication framework (SD-GSC) for efficient image transmission over Rayleigh fading channels. The main goal is to reduce the bandwidth requirement while ensuring high-quality image reconstruction at the receiver. To this end, we first designed a semantic autoencoder to extract semantic information (SI) to reduce the transmitted data size while accurately reconstructing the image from the received SI. To combat the impact of channel noise on the transmitted SI, we designed an SD-based semantic denoiser to better characterize and mitigate the channel noise, leading to significant improvements in both PSNR and FID metrics compared to the state-of-the-art baselines under known channel. 
For unknown channel scenarios, we further designed a parallel SD-GSC (PSD-GSC) to jointly learn the channel gain and denoise the received SI, showcasing the capability of our proposed PSD-GSC in channel estimation and efficient image transmission.
\ifCLASSOPTIONcaptionsoff
  \newpage
\fi

\bibliographystyle{IEEEtran}
\bibliography{ddpm}

\begin{thebibliography}{10}
\providecommand{\url}[1]{#1}
\csname url@samestyle\endcsname
\providecommand{\newblock}{\relax}
\providecommand{\bibinfo}[2]{#2}
\providecommand{\BIBentrySTDinterwordspacing}{\spaceskip=0pt\relax}
\providecommand{\BIBentryALTinterwordstretchfactor}{4}
\providecommand{\BIBentryALTinterwordspacing}{\spaceskip=\fontdimen2\font plus
\BIBentryALTinterwordstretchfactor\fontdimen3\font minus \fontdimen4\font\relax}
\providecommand{\BIBforeignlanguage}[2]{{%
\expandafter\ifx\csname l@#1\endcsname\relax
\typeout{** WARNING: IEEEtran.bst: No hyphenation pattern has been}%
\typeout{** loaded for the language `#1'. Using the pattern for}%
\typeout{** the default language instead.}%
\else
\language=\csname l@#1\endcsname
\fi
#2}}
\providecommand{\BIBdecl}{\relax}
\BIBdecl

\bibitem{Luo_2022_WC}
X.~Luo, H.-H. Chen, and Q.~Guo, ``Semantic communications: Overview, open issues, and future research directions,'' \emph{IEEE Wireless Commun.}, vol.~29, no.~1, pp. 210--219, Jan. 2022.

\bibitem{zhou2022task}
H.~Zhou, Y.~Deng, X.~Liu, N.~Pappas, and A.~Nallanathan, ``Goal-oriented semantic communications for 6g networks,'' \emph{IEEE Internet Things Mag.}, vol.~7, no.~5, pp. 104--110, Aug. 2024.

\bibitem{Bourtsoulatze_2019_TCCN}
E.~Bourtsoulatze, D.~Burth~Kurka, and D.~Gündüz, ``Deep joint source-channel coding for wireless image transmission,'' \emph{IEEE Trans. Cogn. Commun. Netw.}, vol.~5, no.~3, pp. 567--579, May 2019.

\bibitem{Nan2024TWC}
N.~Li, A.~Iosifidis, and Q.~Zhang, ``Dynamic semantic compression for cnn inference in multi-access edge computing: A graph reinforcement learning-based autoencoder,'' \emph{IEEE Trans. Wireless Commun.}, pp. 1--1, Dec. 2024.

\bibitem{Nan2023Globecom}
N.~Li, M.~Bennis, A.~Iosifidis, and Q.~Zhang, ``Spatiotemporal attention-based semantic compression for real-time video recognition,'' in \emph{Proc. IEEE Globecom Workshops (GC Wkshps)}, 2023, pp. 1603--1608.

\bibitem{Xu_2022_TCSVT}
J.~Xu, B.~Ai, W.~Chen, A.~Yang, P.~Sun, and M.~Rodrigues, ``Wireless image transmission using deep source channel coding with attention modules,'' \emph{IEEE Trans. Circ. Syst. Video Tech.}, vol.~32, no.~4, pp. 2315--2328, May 2022.

\bibitem{chung2023diffusion}
H.~Chung, J.~Kim, M.~T. Mccann, M.~L. Klasky, and J.~C. Ye, ``Diffusion posterior sampling for general noisy inverse problems,'' in \emph{Proc. Int. Conf. Learn. Represent. (ICLR)}, Kigali, Rwanda, May 2023.

\bibitem{Rombach_2022_CVPR}
R.~Rombach, A.~Blattmann, D.~Lorenz, P.~Esser, and B.~Ommer, ``High-resolution image synthesis with latent diffusion models,'' in \emph{Proc. IEEE/CVF Comput. Vis. Pattern Recognit. Conf. (CVPR)}, New Orleans, Louisiana, Jun. 2022, pp. 10\,684--10\,695.

\bibitem{Xu_2023_Globecom}
B.~Xu, R.~Meng, Y.~Chen, X.~Xu, C.~Dong, and H.~Sun, ``Latent semantic diffusion-based channel adaptive de-noising semcom for future 6g systems,'' in \emph{Proc. IEEE Global Commun. Conf. (GLOBECOM)}, Kuala Lumpur, Malaysia, Dec. 2023, pp. 1229--1234.

\bibitem{Unet}
O.~Ronneberger, P.~Fischer, and T.~Brox, ``U-net: Convolutional networks for biomedical image segmentation,'' in \emph{Med. Image Comput. Comput. Assist. Interv. (MICCAI)}, Nov. 2015, pp. 234--241.

\bibitem{NEURIPS2020_4c5bcfec}
J.~Ho, A.~Jain, and P.~Abbeel, ``Denoising diffusion probabilistic models,'' in \emph{Adv. Neural Inf. Process. Syst. (NeurIPS)}, vol.~33, Dec. 2020, pp. 6840--6851.

\bibitem{Vincent2011NC}
P.~Vincent, ``A connection between score matching and denoising autoencoders,'' \emph{Neural Comput.}, vol.~23, no.~7, pp. 1661--1674, Jul. 2011.

\bibitem{NEURIPS2021_077b83af}
K.~Kim and J.~C. Ye, ``Noise2score: Tweedie’s approach to self-supervised image denoising without clean images,'' in \emph{Adv. Neural Inf. Process. Syst. (NeurIPS)}, vol.~34, Dec. 2021, pp. 864--874.

\bibitem{Karras_2019_CVPR}
T.~Karras, S.~Laine, and T.~Aila, ``A style-based generator architecture for generative adversarial networks,'' in \emph{Proc. IEEE/CVF Comput. Vis. Pattern Recognit. Conf. (CVPR)}, Jun. 2019, pp. 4396--4405.

\end{thebibliography}

\end{document}